\documentclass[12pt,preprint]{aastex}
\usepackage{graphics}

\shorttitle{XMM-Newton observations of Nova Sgr 1998}
\shortauthors{Hernanz and Sala}

\begin{document}
\title{XMM-Newton observations of Nova Sgr 1998}

\author{M. Hernanz}

\affil{Institut de Ci\`encies de l'Espai (CSIC-IEEC), 
Campus Universitat Aut\`onoma de Barcelona,
Facultat de Ci\`encies, Torre C5 - parell - 2a planta,
E-08193 Bellaterra (Barcelona), Spain; hernanz@ieec.uab.es} 

\and 

\author{G. Sala}
\affil{Max-Planck-Institut f\"ur extraterrestrische Physik,
Postfach 1312, D-85741, Garching, Germany and  
Institut de Ci\`encies de l'Espai (CSIC-IEEC); gsala@mpe.mpg.de}

\begin{abstract}
We report on X-ray observations of Nova Sagittarius 1998 (V4633 
Sgr), performed with XMM-Newton at three different epochs, 934, 1083 and 1265 
days after discovery. The nova was 
detected with the EPIC cameras at all three epochs, with emission 
spanning the whole energy range from 0.2 to 10 keV. 

The X-ray spectra do not change significantly at the different epochs, 
and are well fitted for the first and third observations 
with a multi-temperature optically thin thermal plasma, 
while lower statistics in the second observations lead to a poorer fit. 
The thermal plasma emission is most probably originated in 
the shock heated ejecta, with 
chemical composition similar to that of a CO nova. However, we can not completely 
rule out reestablished accretion as the origin of the emission.
We also obtain upper limits for the temperature and luminosity of a 
potential white dwarf atmospheric component, and conclude that hydrogen burning 
had already turned-off by the time of our observations.
\end{abstract}


\section{Introduction}
Nova Sgr 1998 (V4633 Sgr) was discovered by \cite{Lil98} on 1998 March 22.3~UT, 
with magnitude 7.8, and it was confirmed spectroscopically by \cite{DV98}
two days later, with relatively low expansion velocities and presence of 
iron, thus indicating that the nova belonged to the Fe~II class \citep{Wil92}. 
The nova light curve \citep{LJ99} indicated that it was moderately fast, with 
t$_2\approx$28~days and t$_3\approx$55~days. 
No indication of optical polarization was found in 
further observations by \cite{IKA00}. \cite{Lip01} performed 
optical observations during the period 1998--2000, which revealed two 
photometric periodicities; the shorter and constant period of 3.014~h was 
interpreted as the orbital period, whereas the longer and slightly 
variable period of 3.08~h could be interpreted as the spin period of the 
white dwarf, in a nearly synchronous magnetic system. Near-infrared 
spectroscopy, in the range 0.8 to 2.5~$\mu$m, obtained 525 and 850~days after 
peak brightness by \cite{Lyn01}, revealed some symmetric lines with widths 
(FWHM) of 1800~km/s. There was no evidence of dust formation and the 
shell was hydrogen-deficient. The reddening was very uncertain 
(E(B-V)=$0.3\pm0.2$, yielding A$_V=0.9\pm0.6$ for R=3.1), thus leading to a 
very uncertain distance, d$\sim$9~kpc (d=8.9$\pm$2.5~kpc, \cite{Lip01}). 

Nova Sgr 1998 was one of the targets included in our monitoring programme 
of recent galactic novae in X-rays, during the first cycle 
of the XMM-Newton satellite. The main goal of our observations was to 
determine the turn-off times of hydrogen burning. Novae are the 
consequence of explosive hydrogen burning, through a 
thermonuclear runaway, on top of an accreting white dwarf in a close binary 
system, of the cataclysmic variable type.
It is theoretically predicted that 
novae return to hydrostatic equilibrium after the ejection of a fraction of 
the accreted envelope. X-ray emission in post-outburst novae can arise 
through three different mechanisms. First, as the envelope mass is 
depleted, the photospheric radius decreases at constant bolometric luminosity 
(close to the Eddington value) with an increasing effective temperature. 
This leads to a hardening of the spectrum from optical through 
ultraviolet, extreme ultraviolet and finally soft X-rays, with 
the post-outburst nova emitting as a supersoft source with a 
hot white dwarf atmosphere spectrum.
The duration of this soft X-ray emission is related to 
the nuclear burning timescale of the remaining H-rich envelope 
(see, for instance, Table~1 in \cite{Geh98}) and depends 
among other factors on the white dwarf mass \citep{TT98,SH05a,SH05b}. 
The second site of X-ray emission in post-outburst novae is the ejected shell.
Internal or external shocks can heat the expanding gas up to temperatures of a few keV,
leading to the emission of X-rays with a thermal plasma spectrum. Finally, 
when accretion is reestablished in the cataclysmic variable, 
the accretion flow is responsible for the emission of X-rays, also with 
a thermal plasma spectrum. 

A systematic search for X-ray emission of classical novae
was performed by \cite{Ori01a} in the ROSAT archival data, 
which contained observations of 108 classical and recurrent novae, 
39 of which were less than 10 years old. 
Contrary to expectations, just a few novae were detected in X-rays, and
only three of them with a soft X-ray spectrum: GQ~Mus (Nova~Mus~1983, 
already discovered by EXOSAT in 1983, \cite{Oge84}), 
V1974~Cyg (Nova~Cyg~1992, \cite{Kra96,Bal98}), and 
Nova~LMC~1995 \citep{OG99}. GQ~Mus is renowned for having the longest soft X-ray 
emission phase (around 9 years; \cite{Oge93,Sha95,BK01}).
Other novae were detected in X-rays with ROSAT, but without any 
soft component: V351~Pup (Nova~Pup~1991, \cite{Ori96}), and V838~Her (Nova~Her~1991,
\cite{Llo92,Obr94}). 
Finally, only 11 out of the 81 galactic quiescent novae observed by 
ROSAT were detected in hard X-rays. For all these non soft X-ray sources, 
the poor spectral resolution and/or limited spectral range 
of the detectors left the origin of the emission 
(either shocked ejecta or reestablished accretion) unclear.

After the ROSAT era, observations by Beppo-SAX and Chandra revealed new  
interesting results. A soft component was reported for
V382~Vel (Nova~Vel~1999) which lasted for 7.5--8 months and contained
a wealth of emission lines never previously detected in novae \citep{Ori02,Bur02}.
In the case of V1494~Aql (Nova~Aql~1999~No.2), a soft component appeared 6--10 months after the 
explosion, with a puzzling light curve including a short burst 
and oscillations \citep{Dra03}. Observations of the recurrent nova CI~Aql 
by \cite{Gre02} showed a soft component, but its spectrum implied an emitting 
radius of only 40~km for this source, which is too small for a white dwarf.
In addition, supersoft X-ray emission 6 months after outburst, with strong 
temporal variation was reported by \cite{Nes03} and \cite{Pet05} 
for V4743~Sgr (Nova~Sgr~2002~No.3).

XMM-Newton also contributed with some more observations of novae.
X-ray emission with a soft spectrum was detected from
Nova~LMC~1995 by \cite{Ori03} more than 5.5 years after the explosion.
This component was not detected in later observations by \cite{Ori04} in 2003. 
X-rays were also detected from Nova~LMC~2000 by \cite{Gre03} on days 17 and 51 after outburst, 
with a thermal plasma emission related to the ejecta and fading away 
on day 294. Finally, in the case of V2487~Oph (Nova~Oph~1998), \cite{HS02} 
detected X-ray emission with a broad thermal plasma spectrum extending up to $\sim 10$~keV and
a fluorescent Fe K$_\alpha$ line, clearly indicating the reestablishement of accretion in a 
magnetic cataclysmic variable.

X-ray observations of classical novae during their post-outburst
stages provide crucial information about the nova phenomenon 
\citep{Ori99,Ori04,Kra02}: soft X-rays yield a unique insight 
into the remaining nuclear burning on top of the white dwarf, 
while hard X-rays reflect the shocks in the ejecta in the post-outburst novae, 
and the properties of the accretion flow in the ``quiescent novae''. 
In view of the scarcity of objects observed and the diversity of 
behaviours detected, only the monitoring of as many novae as possible, 
with large sensitivity and spectral resolution (as those offered 
by XMM-Newton and Chandra) can help to understand and solve these issues. 
The observations of Nova Sgr 1998 with XMM-Newton reported in 
this paper contribute to achieve this goal. 
In \S2 we describe the observations, 
and in \S3 the spectral analysis of the data is explained in detail. 
Discussion follows in \S4 and a summary in \S5.

\section{Observations}
The X-Ray Multi-Mirror Mission, XMM-Newton, is an X-ray 
astrophysics observatory (see \cite{Jan01} for a general description)
including three X-ray telescopes, each with an 
European Photon Imaging Camera (EPIC) at its focal plane, and an optical/UV 
telescope, OM \citep{Mas01}. The detectors of two of the EPIC cameras use 
MOS CCDs \citep{Tur01}, 
whereas the third one uses pn CCDs \citep{Str01}. In addition, two 
Reflection Grating Spectrometers \citep{Her01} for high resolution 
X-ray spectroscopy are 
mounted in the light path of the two EPIC MOS cameras.   

We observed Nova Sgr 1998 at three epochs, with intervals of 6 months 
between observations: October 11, 2000, March 9, 2001 
and September 7, 
2001 (i.e., 934, 1083 and 1265 days after outburst; see Table \ref{tab:log}).
Exposure times ranged from 6~ks to 9~ks. 
While the source was clearly detected with the EPIC MOS1, MOS2 and pn cameras, 
the signal to noise ratio was too small to detect it with the RGS 
instruments.

Data were reduced using the XMM-Newton Science Analysis System 
(SAS~6.5.0). Standard procedures described in the SAS documentation 
\citep{sas05,abc04} were applied. 
New calibration files affecting the soft spectrum 
of observations of the first XMM-Newton cycle 
were released in January 2006 \citep{Kir06}, 
and the pipeline products of some exposures of our observations were not available in the
archive. Therefore, we reprocessed the original Observation Data Files (ODFs) 
for all exposures using the most recent calibrations. 
Only events with pattern smaller than 12~(MOS) and 4~(pn) 
and with pulse height (PI) in the range 0.2--12~keV~(MOS) and
0.2--15~keV~(pn) were selected for spectral analysis. 
Flaring background periods affecting some EPIC-pn exposures were 
filtered by creating good time interval (GTI) filter tables that 
were used for further event select procedures. 
Source photons were extracted from a circle of 35'', 
and the background was extracted 
from a region of the same area, close to the source but free from source 
photons, and keeping the same distance to the readout node (RAWY) as the 
source region \citep{Kir06}. The statistics of our observations were not 
good enough to perform timing analysis, which would have been interesting to 
determine the presence of some periodicity, as observed in the optical by \cite{Lip01}.

\section{Spectral analysis}
The X-Ray Spectral Fitting Package (XSPEC~11.3, \cite{xsp03}) was used 
for spectral analysis. For each observation, the spectra of the three EPIC 
cameras (pn, MOS1 and MOS2) were simultaneously fitted. 
An absorbing column density compatible with or smaller than the interstellar 
value was obtained for all spectral models.
A column density smaller than the interstellar N$_{\rm H}$ 
is clearly unphysical, and the spectral fits never indicate a larger value.
Therefore, we froze it in all our fits to N$_{\rm H}$= 1.6$\times 10^{21}$cm$^{-2}$,
which is the average interstellar column density in the direction of Nova~Sgr~1998
\citep{DL90}. This value is also consistent with the N$_{\rm H}$ derived from 
the measured extinction \citep{Lyn01}, through the empirical relationship between 
interstellar X-ray absorption and optical extinction \citep{Gor75}.

\subsection{Spectral models}

A first look at the data (Fig.~\ref{fig:spectra})
clearly shows that the emission spans the whole energy range 
of the EPIC instruments, $0.2-10$~keV, with a spectrum harder than 
that produced by residual H-burning on top of the white dwarf. 
This immediately indicates that the emission must be dominated by either the 
hot ejected shell, or by a reestablished accretion flow.
We thus fit the EPIC spectra with a multi-temperature 
optically thin thermal plasma, simulated with a 
Raymond-Smith model \citep{RS77} in XSPEC. 
The main processes included in this model are bremsstrahlung, 
recombination continua and line emission. More sophisticated models 
such as MEKAL \citep{MKL95} are senseless for our data, since these 
models require a larger number of parameters whereas we have a low number 
of counts. We checked, however, that the results obtained with MEKAL and 
Raymond-Smith models are similar. 
A single temperature thermal plasma model is completely unable to fit the data, 
because the emission spans a broad energy range. 
At least two components at different temperatures are needed to obtain a reasonable fit,
with the best fit being obtained for a three-temperature model.

Since the thermal plasma emission can arise either in the ejecta or in the 
accretion flow, two kinds of chemical composition are possible:
solar abundances (for an accretion flow), or metal enriched (for 
the nova ejecta). In this latter case, the underlying white dwarf of the nova can have 
two different compositions: carbon-oxygen (CO) for masses $\le 1.15$M$_\odot$, 
or oxygen-neon (ONe) for masses up to the Chandrasekhar limit. 
Here we use realistic compositions of the ejecta from \cite{JH98}. 
We tested their 14 ejecta compositions, 
corresponding to CO and ONe nova models with masses ranging 
from 0.8 to 1.15 M$_\odot$ and 1.0 to 1.35 M$_\odot$, respectively, and 
degrees of mixing between accreted mass and underlying core 
from 25 to 75 \%. In general, the best fit is obtained for 
low-mass CO novae (see below for details).

We also considered the possible contribution of residual H-burning on the 
post-outburst white dwarf (again CO or ONe).
We used white dwarf atmosphere emission models, gently 
provided by Jim MacDonald \citep{MV91}, and 
built tables to be read as external models in XSPEC.
Previous studies have shown the importance of including 
white dwarf atmosphere models instead of simple blackbodies 
for a correct spectral analysis of 
novae or supersoft X-ray sources \citep{Bal98,BK01,OG99}.

\subsection{Results}

A three temperature (3T) optically thin thermal plasma model provides a good fit for the 
three EPIC cameras spectra of the first observation. The fit is only slightly poorer
for the third observation, while the poor statistics lead to a much worse 
fit at the second epoch. However, no other model provides a better fit and
the physical origin of the emission is very likely to be the same 
in all our observations. Therefore, we fit the spectra at the three epochs with 
the same 3T plasma model.

\begin{figure}
\begin{center}
\includegraphics[width=0.7\textwidth]{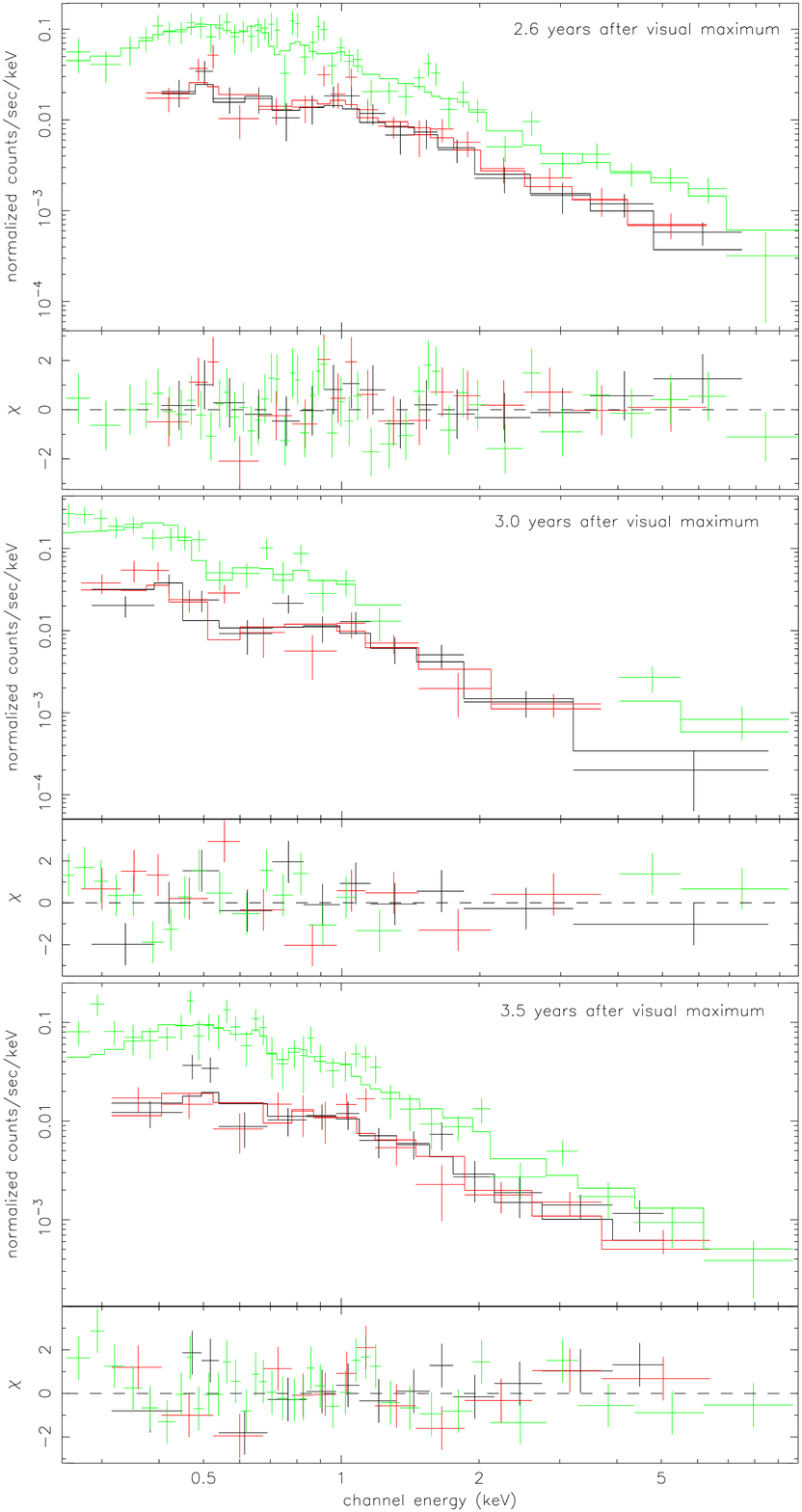}
\caption{Nova Sgr 1998 EPIC MOS1 (black), MOS2 (red) and pn (green) observed spectra, with the best fit 
model (with CO abundances) and the residuals. From top to low panels: first,
second and third observations. 
\label{fig:spectra}}
\end{center}
\end{figure}

Figure~\ref{fig:spectra} (upper panel) shows Nova~Sgr~1998 spectra in October 2000, 
934~days (2.6~years) after the explosion, with the best fit model and the 
residuals in units of $\chi$. 
The 3T plasma model provides a good fit to the EPIC MOS1, MOS2 and pn data, 
both simultaneously and individually.
The 3T model was used only after checking 
that a two temperature plasma model (which would be preferred because of its simplicity) 
results in a worse fit (see Table~\ref{tab:models_2T}), 
leaving significant residuals, in particular around 1~keV. 
We also 
attempted to fit the spectra with a cooling flow model, 
which is a good representation of a multi-temperature plasma in 
accretion disks in old novae \citep{Muketal03}. However, the spectrum is not well 
fitted with this model: with T$_{\rm low}$=0.08~keV, T$_{\rm high}$=3~keV 
and normalization constant $10^{-9}$M$_\odot$/yr, the reduced $\chi^2$ is 2.9;
therefore ºwe should rule out the cooling flow model 
for Nova~Sgr~1998.

\begin{figure}
\begin{center}
\includegraphics[width=0.5\textwidth]{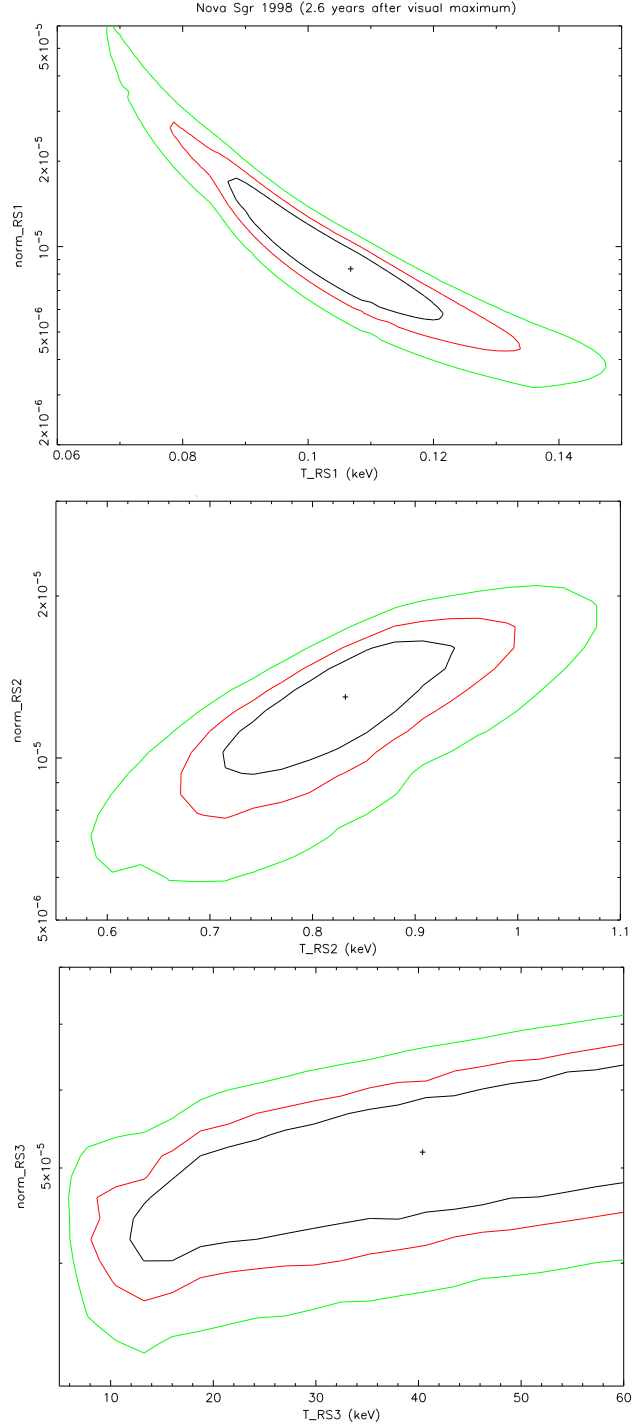}
\caption{Confidence contours (1-innermost-, 2, and 3 -outermost- $\sigma$)
for the temperature and normalization constant 
($\rm norm_{RS}=10^{-14}$EM$/(4\pi$D$^{2})$, 
where EM is the emission measure and D the distance in cm)
of the low (upper panel), intermediate (middle panel) and high (lower panel) 
temperature thermal plasma components, for the models with
CO abundances corresponding to the first observation of Nova Sgr 1998. 
\label{fig:contorns}}
\end{center}
\end{figure}

Table~\ref{tab:models_3T} shows the 
best-fit parameters for the 3T plasma models both with solar and  
with CO nova abundances. In both cases, the three temperatures are
around 0.1, 0.8 and $> 5$ keV (probably indicating a continuous distribution 
between the two extremes). The main difference between the results with 
different abundances lies in the emission measures 
(EM, defined as ${\rm \int n_e n_i dV \simeq \int n_e^2 dV}$, with 
n$_{\rm e}$ and n$_{\rm i}$ the electronic and ionic densities, respectively).
The EM for the coolest component in the solar case is extremely large (by itself and 
when compared with the EM of the intermediate and high temperature components). 
But this problem disappears when a CO plasma is considered, because the emission 
measure for the coolest component decreases by a factor of 100. 
The difference is due to the strong emission 
lines at energies at 0.1-1~keV for the C, N and O enriched abundances. 
Confidence contours for the parameters of the three components are displayed 
in Figure~\ref{fig:contorns}.

%The corresponding luminosities are 
%L${\rm _{unabs.}}$ (0.2-10.0 keV) = (3-13)$\times 10^{33}\times ({\rm d/9kpc})^2$ 
%erg/s and L${\rm _{unabs.}}$ (0.2-10.0 keV) = 
%(2-8)$\times 10^{33}\times ({\rm d/9kpc})^2$ erg/s, for solar and CO cases, 
%respectively.

\begin{figure}
\begin{center}
\includegraphics[width=0.8\textwidth]{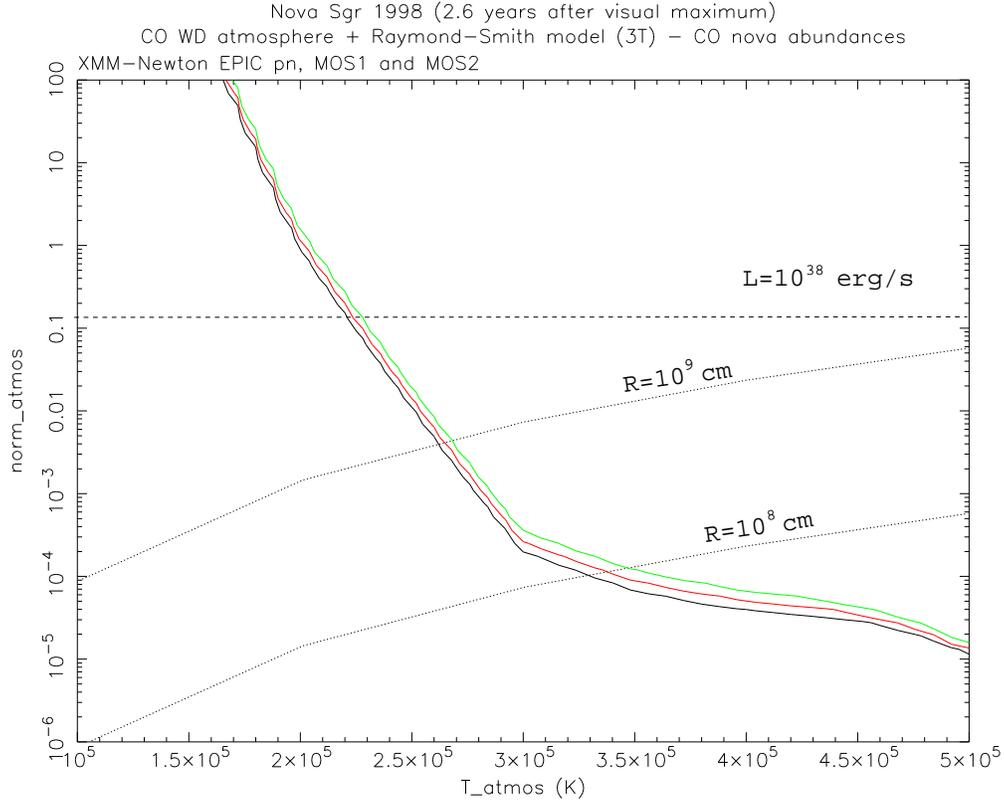}
%{sgr98_1a_contatmos_atm3rs_co1_paper.ps}
%\includegraphics[totalheight=6.0cm]
\caption{Upper limits
(from left to right, 1, 2, and 3 $\sigma$ confidence) 
for the effective temperature (T$_{\rm atmos}$) and normalization
(norm=L$_{39}$/D$^2_{10}$, where L$_{39}$ is the luminosity in units of 
$10^{39}$erg/s and D$_{10}$ the 
distance to the source in units of 10 kpc) of the CO white dwarf atmosphere 
component (first observation). The thermal plasma model has three components 
and abundances corresponding to CO novae ejecta, as in previous figures.
The luminosity of the nova during the constant bolometric luminosity phase, 
L$\sim$$10^{38}$ erg/s (dashed line), as well as the locii of constant 
white dwarf radius equal to $10^8$ and $10^9$ cm (dotted lines) are 
also displayed.
\label{fig:contatm}}
\end{center}
\end{figure}

The presence of a soft X-ray atmospheric component in the spectrum of Nova Sgr 1998 
can not be ruled out, but only upper limits can be established (Fig.~\ref{fig:contatm}).
They include a broad confidence region of the T$_{\rm eff}$-L$_{\rm bol}$ parameters,
but taking into account the path of a post-outburst nova on the T$_{\rm eff}$-L$_{\rm bol}$ diagram, 
we can conclude that hydrogen burning had already turned off by the time of our first observation.
The initial evolution of a post-outburst nova follows a constant bolometric luminosity path  
(close to Eddington limit, i.e., $\sim 10^{38}$erg/s for a 1M$_{\odot}$ white dwarf) with increasing 
effective temperature \citep{KH94,SH05b}.
If Nova Sgr 1998 had been in this first phase with high luminosity 
at the time of our observation, 2.7~years after the outburst, 
it would have been at high temperature. 
But for a high luminosity, the upper limits for our first observation 
are compatible only with low temperatures, which would indicate an extended 
envelope. This is highly unlikely for a nova 2.6~years after the outburst.
Therefore we conclude that hydrogen burning had already turned-off 
by the time of our first observation.
The upper limits are however compatible with a cooling white dwarf: 
taking $10^8$cm as a hard lower limit for the radius of the degenerate star,  
Fig.~\ref{fig:contatm} indicates that it was colder than $3\times10^5$K. 

The second observation (3.0~years after the explosion, 
middle panel in Fig~\ref{fig:spectra}) provided lower 
signal to noise data (due to a smaller exposure time, 
$\sim 30\%$ less than in the first observation, Table~\ref{tab:log}). 
For completeness, we include a fit similar to that used for the first and third observations. 
Despite the large error ranges, the best-fit parameters  
(Table \ref{tab:models_3T}) indicate a slight softening of the spectrum, 
with lower temperatures for the first and third component.

The fit results for the third observation (3.5~years after the explosion, 
lower panel in Fig~\ref{fig:spectra}), 
are similar to the first one. However, 
the emission measures for the intermediate and highest 
temperature components are a bit smaller, whereas that associated with the coolest 
component is larger than one year before. This evolution is compatible with some 
cooling of the plasma between the first and third epochs. 
 
%The unabsorbed flux (in the observed range 0.2-10.0 keV) is 
%(2-33)$\times 10^{-13}$photons/cm$^2$/s (solar) and 
%(2-6)$\times 10^{-13}$photons/cm$^2$/s (CO) (see Tables \ref{tab:models_solar} 
%and \ref{tab:models_co1}). 
%The corresponding luminosities are 
%L${\rm _{unabs.}}$ (0.2-10.0 keV) = (2-32)$\times 10^{33}\times ({\rm d/9kpc})^2$ 
%erg/s and L${\rm _{unabs.}}$ (0.2-10.0 keV) = 
%(2-6)$\times 10^{33}\times ({\rm d/9kpc})^2$ erg/s, for solar and CO cases, 
%respectively.

\section{Discussion}

The optically thin thermal plasma emission may originate either 
in the shocked expanding nova shell or in the accretion flow. 
The problem of distinguishing between these two possibilities has already
been faced in the case of other sources:
Nova~Pup~1991 \citep{Ori96}, Nova~Her~1991 \citep{Llo92,Obr94}, 
Nova~Cyg~1992 \citep{Kra96,Bal98}, Nova~Vel~1999 \citep{Ori01b,MI01}. 
We discuss below some properties of the models which could help to disentangle the 
origin of the emission: luminosity and emission measure (i.e., value and 
temporal evolution of electronic density).

The unabsorbed X-ray (0.2--10 keV) luminosity for Nova~Sgr~1998
is in the range $10^{33}-10^{34}$~erg/s in all cases, for a distance of 9~kpc \citep{Lip01}. 
Distances to novae are quite uncertain, and Nova~Sgr~1998 is 
not an exception (d=8.9$\pm$2.5~kpc), but a factor of $\le 2$ in distance only 
changes the luminosity by a factor of $\le 4$.
This luminosity is too large to originate from the accretion
onto a non-magnetic white dwarf (typically $10^{29}-10^{32}$~erg/s, \cite{Kuu03}), 
or in a polar cataclysmic variable (with luminosities $\sim10^{32}$~erg/s, \cite{Muk03,Ram04}).
However, it could be related to accretion in an intermediate polar, 
with luminosities $\sim10^{33}$~erg/s \citep{Muk03}. In fact, optical observations suggested 
that Nova~Sgr~1998 is a nearly synchronous
magnetic system \citep{Lip01}, and photometric
observations of other post-novae indicate that
some are intermediate polars (for instance, V697~Sco, \cite{WW02}).

The X-ray luminosity of Nova~Sgr~1998 is also compatible with 
the observed values for nova ejecta. For example, 
Nova~Her~1991 emitted $5\times10^{33}-10^{35}$~erg/s only five days after outburst,
fading afterwards extremely quickly \citep{Llo92,Obr94};
Nova~Cyg~1992 ejecta X-ray emission reached $(0.8-2.0)\times 10^{34}$~erg/s 
around 150~days after outburst \citep{Bal98}; 
Nova~Pup~1991 emitted around $7.5\times 10^{33}$~erg/s 16~months 
after outburst \citep{Ori96}; and Nova~Vel~1999 also reached  
L$>10^{34}$~erg/s in the band above 0.8~keV 15~days after maximum 
(the shell luminosity was even larger, since emission lines detected below 0.8~keV 
also originated in the ejecta),
but it decreased to a few $10^{33}$~erg/s 5.5~months later \citep{Ori01b,Ori02}. 
In all those previous cases the ejecta cooled much faster than in Nova Sgr 1998. 
However, the expected cooling time for the electronic densities, n$_{\rm e}$, in nova ejecta 
(in general smaller than $10^6\rm{cm}^{-3}$) are larger than 10~years \citep{Bre77,Oge87}. 
Therefore, X-ray emission from shock heated ejecta in Nova~Sgr~1998 3.5~years after the 
outburst would be fully compatible with theoretical predictions of cooling times.

For a shock-heated plasma, temperatures are related to velocity as kT$\propto\rm{v}^{2}$
\citep{Bre77}. For the expansion velocities observed by \cite{Lyn01} 
(1800~km/s, on days 525 and 850 after outburst),
the plasma temperature would be $\sim$ 3 keV, which lies
within the range of temperatures of our best-fit thermal plasma models. 
Coronal lines, which are a good signature of shocks in the plasma, 
were also present in those observations, supporting the presence of a 
shocked ejecta in Nova~Sgr~1998. 
Assuming an expansion at constant velocity,
the emission volume at the time of our observations
would be increasing from $1.3\times 10^{49}\,\rm{cm}^{3}$ (for day~934) 
to $3.2\times 10^{49}\,\rm{cm}^{3}$ (day 1265).
With these volumes and the emission measures for CO models 
from Table~\ref{tab:models_3T}, electronic densities, $\rm{n}_{\rm{e}}$, in the shell would 
evolve from $1.5\times10^3\,\rm{cm}^{-3}$ (day~934) 
to $1.3\times10^3\,\rm{cm}^{-3}$ (day~1265) for the coolest component. Similarly,
$\rm{n}_{\rm{e}}$ decreases from $10^3\,\rm{cm}^{-3}$ to 500~cm$^{-3}$, and 
from $2\times10^3\,\rm{cm}^{-3}$ to $10^3\,\rm{cm}^{-3}$ 
for the intermediate and high temperature components, respectively. 
Since the spherical geometry gives the maximum emitting volume, these $\rm{n}_{\rm{e}}$
are lower limits. The small $\rm{n}_{\rm{e}}$ agree with the 
thinning of the ejecta found between days 525 and 850 by \cite{Lyn01}.

The electronic density in a spherical shell with its outer radius expanding 
at constant velocity evolves with time as t$^{-3}$, while in an expanding shell 
of constant thickness $\rm{n}_{\rm{e}}$ evolves as t$^{-2}$ \citep{Lyk03}. 
With the first expansion picture, a decrease of $\rm{n}_{\rm{e}}$ 
by a factor $\sim2.5=1/0.4$ is expected between
the first and the third observations, while the decay would
be by a factor of $\sim1.8=1/0.5$ in the second case. 
For Nova Sgr 1998, the $\rm{n}_{\rm{e}}$ derived above for the CO model 
changes by a factor of 0.8 for the coolest thermal plasma component.
For the intermediate and hottest components $\rm{n}_{\rm{e}}$ decreases 
by 0.5, the expected factor for an expanding shell of constant thickness.
Therefore, the $\rm{n}_{\rm{e}}$ evolution also supports the expanding ejecta as the 
possible site of the observed X-ray emission.
However, the large errors associated with n$_{\rm e}$ 
(in particular for the lowest temperature component) 
prevent a clear conclusion about the shell geometry.

\section{Summary}

The X-ray spectrum of Nova~Sgr~1998 as observed by XMM-Newton between 2.6 and 3.5 years 
after outburst is dominated by thermal plasma emission.
The best-fit is obtained with a three-component thermal plasma model, with temperatures
between 0.1 and 40~keV. Thermal plasma models with
different compositions (solar, and several abundances from  
realistic nova ejecta) have been tried. The best-fit is obtained with either the abundances of a 
CO nova shell (for 0.8~M$_\odot$ and 25\% mixing), 
or with solar abundances. 
The first case would correspond to emission from the expanding nova shell, 
while the second would be associated with reestablished accretion 
in the cataclysmic variable. 

The X-ray luminosity, the values of the plasma temperatures, 
the chemical composition, and the distribution and 
evolution of emission measures point to shock heated ejecta as the most likely 
origin for the plasma emission, 
rather than resumed accretion in the cataclysmic variable.
Also the inability to fit the data with a cooling flow model casts doubts on  
accretion as the origin of the X-ray emission.
Finally, the spectra of Nova~Sgr~1998 do
not show any fluorescence iron line, contrary to what was observed for
V2478~Oph \citep{HS02}, which clearly indicated that accretion 
was reestablished in the cataclysmic variable.

Regarding the presence of residual hydrogen burning on the white dwarf surface, 
our observations indicate that the post-outburst nova envelope had already 
turned off its hydrogen burning already 2.6~years after optical maximum.

\acknowledgments
We thank Jim MacDonald for kindly providing his white dwarf atmosphere 
models, and the referee for many helpful comments, which have considerably
improved this paper.
This work is based on observations obtained with XMM-Newton, an ESA science
mission with instruments and contributions directly funded by ESA Member
States and NASA. 
The XMM-Newton project is supported by the
Bundesministerium f\"ur Wirtschaft und Technologie/Deutsches Zentrum
f\"ur Luft- und Raumfahrt (BMWI/DLR, FKZ 50 OX 0001), the Max-Planck
Society and the Heidenhain-Stiftung.
This research has made use of the SIMBAD database,
operated at CDS, Strasbourg, France, and has been partially funded by the 
Spanish Ministry for Education and Science (MEC) project AYA2004-06290-C02-01 
and by FEDER. G.S. is supported through a postdoctoral fellowship from the 
MEC.

\clearpage

% Tables

\begin{deluxetable}{rrccrccrcc} 
\tabletypesize{\scriptsize}
\tablecolumns{10} 
\tablewidth{0pt} 
\tablecaption{Observation log.
\label{tab:log}}
\tablehead{ 
\colhead{} & \colhead{} & 
\multicolumn{2}{c}{EPIC pn} & \colhead{} & 
\multicolumn{2}{c}{EPIC MOS1} & \colhead{} &
\multicolumn{2}{c}{EPIC MOS2} \\ 
\cline{3-4} \cline{6-7} \cline{9-10} \\ 
\colhead{Observation}  & \colhead{Time after}  & 
\colhead{Exposure}     & \colhead{Count Rate} & \colhead {} &
\colhead{Exposure}     & \colhead{Count Rate} & \colhead {} &
\colhead{Exposure}     & \colhead{Count Rate}\\ 
\colhead{Date (orbit)} & \colhead{Discovery}&
\colhead{Time (s)}     & \colhead{($10^{-2}\rm cts$ $\rm s^{-1}$)} 
                       & \colhead {} &
\colhead{Time (s)}     & \colhead{($10^{-2}\rm cts$ $\rm s^{-1}$)} 
                       & \colhead {} &
\colhead{Time (s)}     & \colhead{($10^{-2}\rm cts$ $\rm s^{-1}$)}}
\startdata
Oct.11, 2000 (154) & 934 d, 2.6 yr&
6213 & $10.1\pm0.5$ & &
9048 & $2.4\pm0.2$ & &
9051 & $2.7\pm0.2$ \\ 

Mar. 9, 2001 (229) & 1083 d, 3.0 yr&
3058 & $8.6\pm0.6$ & &
6575 & $2.1\pm0.2$ & &
6575 & $2.2\pm0.2$ \\ 

Sep. 7, 2001 (320) & 1265 d, 3.5 yr&
5883 & $8.2\pm0.4$ & &
9086 & $2.0\pm0.2$ & &
9086 & $1.9\pm0.2$ \\

\enddata
\end{deluxetable} 

\begin{deluxetable}{lccc}
\tabletypesize{\scriptsize}
\tablecolumns{4} 
\tablewidth{0pt} 
\tablecaption{Parameters of the fit for each one of the three 
observations of Nova Sgr 1998, with a two temperatures thermal plasma 
(Raymond-Smith), with solar and CO1 abundances.
\label{tab:models_2T}}
\tablehead{ 
\colhead{} & \colhead{$1^{\rm st}$ Observation} & 
\colhead{$2^{\rm nd}$ Observation} & \colhead{$3^{\rm rd}$ Observation}\\
\cline{1-4}\\
\multicolumn{4}{c}{Solar abundances}}
\startdata
${\rm kT_{RS1} (keV)}$  
                        & $0.12-0.20$
                        & $0.03-0.06$
                        & $0.06-0.12$\\

%${\rm K_{1}}$
%                       & $(0.7-2.8)\times 10^{-4}$
%                       & $0.02-2$
%                       & $(3-118)\times 10^{-4}$ \\

${\rm EM_{RS1} (\times 10^{57} cm^{-3})}$ 
                        & $0.07-0.27$ 
                        & $20-2000$ 
                        & $0.3-11.3$\\

${\rm kT_{RS2} (keV)}$  
                        & $4-11$
                        & $2-16$
                        & $2-6$   \\

%${\rm K_{2}}$
%                       & $(1-2)\times 10^{-4}$
%                       & $(0.6-1.0)\times 10^{-4}$
%                       & $(0.8-1.2)\times 10^{-4}$ \\

${\rm EM_{RS2} (\times 10^{55} cm^{-3})}$  
                        & $12-15$
                        & $6-10$
                        & $8-12$\\

F$_{\rm unabs,0.2-10.0\, \rm keV} (\times 10^{-13} \rm erg\, cm^{-2}\, s^{-1})$ 
                        & $4-7$ 
                        & $2-210$ 
                        & $2-30$ \\

L$_{\rm unabs,0.2-10.0\, \rm keV} (\times 10^{33} \rm erg\, s^{-1})$ 
                        & $4-7$ 
                        & $2-203$ 
                        & $2-29$ \\

$\chi^{2}_{\nu}$        
                        & 1.55
                        & 2.07
                        & 1.66 \\

\cutinhead{CO abundances}

${\rm kT_{RS1} (keV)}$  
                        & $0.09-0.19$ 
                        & $0.03-0.07$ 
                        & $0.07-0.15$ \\

%${\rm K_{1}}$
%                       & $(0.3-1.8)\times 10^{-5}$
%                       & $10^{-4}-0.3$
%                       & $(0.4-4.5)\times 10^{-5}$ \\

${\rm EM_{RS1} (\times 10^{55} cm^{-3})}$ 
                        & $0.3-1.7$ 
                        & $10-3\times10^{4}$ 
                        & $0.4-4.4$  \\

${\rm kT_{RS2} (keV)}$  
                        & $3-8$ 
                        & $1-6$ 
                        & $2-6$    \\

%${\rm K_{2}}$
%                       & $(5-6)\times 10^{-5}$ 
%                       & $(1-4)\times 10^{-5}$ 
%                       & $(3-5)\times 10^{-5}$ \\

${\rm EM_{RS2} (\times 10^{55} cm^{-3})}$  
                        & $5-6$
                        & $1-4$
                        & $2-5$   \\

F$_{\rm unabs,0.2-10.0\, \rm keV} (\times 10^{-13} \rm erg\, cm^{-2}\, s^{-1})$ 
                        & $4-6$ 
                        & $1-80$ 
                        & $2-6$ \\

L$_{\rm unabs,0.2-10.0\, \rm keV} (\times 10^{33} \rm erg\, s^{-1})$ 
                        & $4-6$ 
                        & $1-77$ 
                        & $2-6$ \\

$\chi^{2}_{\nu}$        
                        & 1.37
                        & 2.01
                        & 1.57 \\
\enddata
\tablecomments{The best-fit models shown fit simultaneously the data 
from the three EPIC cameras.
All the limits are $3\sigma$.
The emission measures and luminosities are 
given for a distance d=9kpc; a factor 
${\rm (d/9 kpc)^2}$, with d the distance in kpc, affects these magnitudes.
The absorption ${\rm N_H}$ is $1.6 \times 10^{21} {\rm cm^{-2}}$ everywhere.}
\end{deluxetable}

\begin{deluxetable}{lccc}
\tabletypesize{\scriptsize}
\tablecolumns{4} 
\tablewidth{0pt} 
\tablecaption{Parameters of the best-fit model for each one of the three 
observations of Nova Sgr 1998: three temperatures thermal plasma 
(Raymond-Smith), with solar and CO1 abundances. 
\label{tab:models_3T}}
\tablehead{ 
\colhead{} & \colhead{$1^{\rm st}$ Observation} & 
\colhead{$2^{\rm nd}$ Observation} & \colhead{$3^{\rm rd}$ Observation}\\
\cline{1-4}\\
\multicolumn{4}{c}{Solar abundances}}
\startdata

${\rm kT_{RS1} (keV)}$  
                        & $0.07-0.16$  
                        & $0.03-0.05$
                        & $0.06-0.10$\\

${\rm EM_{RS1} (\times 10^{57} cm^{-3})}$ 
                        & $0.1-2.4$ 
                        & $50-5000$ 
                        & $0.5-6.4$\\

${\rm kT_{RS2} (keV)}$  
                        & $0.6-1.0$
                        & $0.2-0.7$
                        & $0.2-1.0$   \\

${\rm EM_{RS2} (\times 10^{55} cm^{-3})}$  
                        & $0.9-3.7$
                        & $0.5-3.9$
                        & $0.4-2.0$\\

${\rm kT_{RS3} (keV)}$  
                        & $\geq5$
                        & $\geq2$
                        & $\geq3$   \\

${\rm EM_{RS3} (\times 10^{55} cm^{-3})}$  
                        & $8-14$
                        & $5-10$
                        & $6-10$\\

F$_{\rm unabs,0.2-10.0\, \rm keV} (\times 10^{-13} \rm erg\, cm^{-2}\, s^{-1})$ 
                        & $3-13$ 
                        & $2-640$ 
                        & $2-33$ \\

L$_{\rm unabs,0.2-10.0\, \rm keV} (\times 10^{33} \rm erg\, s^{-1})$ 
                        & $3-13$ 
                        & $2-620$ 
                        & $2-32$ \\

$\chi^{2}_{\nu}$        
                        & 1.16
                        & 1.40
                        & 1.22 \\

\cutinhead{CO abundances}

${\rm kT_{RS1} (keV)}$  
                        & $0.07-0.15$
                        & $0.02-0.07$
                        & $0.05-0.13$\\

${\rm EM_{RS1} (\times 10^{55} cm^{-3})}$ 
                        & $0.3-5.8$ 
                        & $14-500000$ 
                        & $0.5-11.0$\\

${\rm kT_{RS2} (keV)}$  
                        & $0.6-1.1$
                        & $0.4-1.2$
                        & $0.4-1.1$   \\

${\rm EM_{RS2} (\times 10^{55} cm^{-3})}$  
                        & $0.6-2.0$
                        & $0.4-1.9$
                        & $0.2-1.7$\\

${\rm kT_{RS3} (keV)}$  
                        & $\geq5$
                        & $\geq2$
                        & $\geq3$   \\

${\rm EM_{RS3} (\times 10^{55} cm^{-3})}$  
                        & $3-7$
                        & $1-5$
                        & $2-5$\\

F$_{\rm unabs,0.2-10.0\, \rm keV} (\times 10^{-13} \rm erg\, cm^{-2}\, s^{-1})$ 
                        & $2-8$ 
                        & $1-5500$ 
                        & $2-6$ \\

L$_{\rm unabs,0.2-10.0\, \rm keV} (\times 10^{33} \rm erg\, s^{-1})$ 
                        & $2-8$ 
                        & $1-5300$ 
                        & $2-6$ \\

$\chi^{2}_{\nu}$        
                        & 0.92
                        & 1.61
                        & 1.25 \\
\enddata 
\tablecomments{Same comments as in table \ref{tab:models_2T}.}
\end{deluxetable}

%% End of tables

\clearpage

%% Figures

\end{document}